%Paper: hep-th/9205068
%From: vadim@bolvan.ph.utexas.edu (Vadim S. Kaplunovsky)
%Date: Tue, 19 May 92 17:33:43 CDT

%&phyzzx
%macropackage=phyzzx
%
% If you do not have the phyzzx.fmt format,
% uncomment the following line:
%\input phyzzx
\hsize=6.5in
%\vsize=9in
%\nopagenumbers
\def\str{\mathop{\rm str}\nolimits}

\centerline{\it ERRATA}\smallskip
\centerline{\fourteenrm One-Loop Threshold Effects in String
Unification%
\foot{Originally published in \sl Nucl.~Phys.~\bf B307 \rm
(1988),pp.~145--156}}
\smallskip
\centerline{Vadim S.\ Kaplunovsky\foot{%
Current research supported in part by the NSF under grant
\#PHY--90--09850
and by the Robert~A.\ Welch Foundation.  Research leading to the
original publication was supported in part by the NSF under grant
\#PHY--86--12280}}
\centerline{\it Department of Physics, University of Texas,
Austin, TX~78712, USA\rm\foot{Current address.}}
\bigskip

As originally published, the article contained a few minor
errors.
Unfortunately, these errors have lately caused a not-so-minor
confusion among several physicists working on the phenomenology
of string unification.
To avoid further confusion, I would like to make the following
corrections:

\noindent {\bf(1)}\enspace
The renormalization scheme referred throughout the article as
$\overline{\rm MS}$ is in fact the modified minimal subtraction
scheme for the {\sl dimensional reduction} and not the standard
dimensional regularization.
The proper name for the scheme I used is $\overline{\rm DR}$.

\noindent {\bf(2)}\enspace
In the second paragraph on page~154 the formula for $\xi''$ should
be\hskip 0pt plus 1in\strut\
${\xi''=1+\log(2/\sqrt{27}\pi)-\gamma}\approx -1.6767$ instead of
$\xi''=1-\log(2/\sqrt{27}\pi)-\gamma\approx -0.532$.

The rest of the mistakes involve misplaced factors of 2:\brk
{\bf(3)}\enspace
Formula (26) for the effective scale of string unification should read
$$
M_{\rm GUT}\ \buildrel \rm def\over= \ {{\bf 2}e^{(1-\gamma)/2}3^{-3/4}
\over \sqrt{2\pi\alpha'}}\
\approx\ {e^{(1-\gamma)/2}3^{-3/4}\over 4\pi}\,g_{\rm string}M_{\rm
Planck}\
\approx\ g_{\rm GUT}\times 5.27\cdot 10^{17}\,\rm GeV.
\eqno(26)
$$
(In the original publication, the factor {\bf 2} shown here in the
boldface
was missing.)
Note that the correct formula here uses the tree-level relation
$k_ag_a^2=g^2_{\rm GUT}=32\pi/\alpha' M^2_{\rm Planck}$ which
differs by a factor of 2 from a similar formula given in ref.~[6].
This difference is due to different normalization conventions
for the gauge generators $Q_a$ and gauge couplings $g_a$ ---
unlike Ginsparg, I used the phenomcnological convention\foot{%
    Most phenomenologists normalize the non-abelian
    gauge generators to $\tr(Q_a^2)=\half$ where the trace is taken over
    the fundamental representation of an $SU(N)$ group such as
    $SU(3)_{\rm color}$ or $SU(2)_{\rm weak}$; correspondingly,
    the generators of a GUT group are normalized to $\tr_{\bf
5}(Q_a^2)=\half$
    for the $SU(5)$ or $\tr_{\bf 10}(Q_a^2)=1$ for the $SO(10)$.
    On the other hand, many string theorists normalize the generators to
    $\tr(Q_a^2)=2$ where the trace is taken over a vector representation
    of an $SO(2N)$ group such as $SO(32)$ or $SO(16)\subset E_8$;
    according to this convention, phenomenologically-normalized $Q_a$
    should be multiplied by $\sqrt{2}$.
    In both conventions, the gauge connection $A_\mu$ is $ig Q_a
A^a_\mu$,
    $A^a_\mu$ being canonically-normalized vector fields;
    therefore, the string-theoretical convention has to compensate
    for the $Q_a$ being bigger by a factor $\sqrt{2}$ by having the
    gauge couplings $g$ being smaller by the same factor.}
throughout the paper and adjusted the string-theoretical formul\ae\
to work with this convention.

\noindent {\bf(4)}\enspace
In formul\ae\ (1), (2), (7) and (24) coefficients printed as $4\pi^2$
should be $16\pi^2$.
Similarly, in formul\ae (5) and (21) coefficients in front of the
respective
integrals should be $1/16\pi^2$ instead of $1/4\pi^2$ and in
formula~(16)
the coefficient printed as $2{\alpha'}^2$ should be ${\alpha'}^2/2$.

\noindent {\bf(5)}\enspace
In formul\ae~(6) and (23), the expressions for
the coefficients $b_a$ of the low-energy $\beta$-functions were too
small
by a factor of 2.
Correspondingly, the field-theoretical functions ${\bf B}_a(t)$ and
the string-theoretical functions ${\cal B}_a(\tau,\bar\tau)$ also
missed that factor.
The correct form of the field-theoretical eq.~(5) is
$$
{\cal W}^{\rm field}_a\
=\ {1\over16\pi^2} \int_0^\infty {dt\over t} C_\Lambda(t)\cdot
\left[ {\bf B}_a(t) \buildrel \rm def\over= {\bf2}\str\left(
    Q_a^2\bigl(\coeff{1}{12}-\chi^2\bigr) e^{-tM^2}\right) \right] .
\eqno(5)
$$
Consequently, the super-traces in eqs.~(7) and (8) should also be
multiplied
by two; in particular, the correct formula for the
one-loop threshold corrections in GUTs is
$$
\Delta_a\ =\ {\bf2}\str_{M\sim M_{\rm GUT}}\left( Q_a^2
    \bigl(\coeff{1}{12}-\chi^2\bigr)\,\log{M^2_{\rm GUT}\over M^2}
\right) .
\eqno(8)
$$

\noindent
Similarly, the correct form of the string-theoretical formula~(22) is
$$
{\cal B}_a(\tau,\bar\tau)\
=\ {{\bf2}\over|\eta(\tau)|^4} \sum_{\rm even\ \bf s} (-)^{s_1+s_2}\,
{d Z_\Psi(\bar\tau,{\bf s})\over 2\pi i\,d\bar\tau}\cdot
\Tr_{s_1}\left( Q_a^2\cdot (-)^{s_2F} q^H {\bar q}^{\bar H}\right)_{\rm
int}\,.
\eqno(22)
$$
The relation (25) between these ${\cal B}_a$ functions and the

string threshold corrections $\Delta_a$ remains unchanged,
which means that the actual values of $\Delta_a$ should be doubled.

%For phenomenological purposes, this is the only non-obvious mistake
%that really matters.

\noindent
As to the formul\ae~(6) and (23) for $b_a$ and the infra-red limits of
${\bf B}_a(t)$ and ${\cal B}_a(\tau,\bar\tau)$, in addition to missing
an overall factor of 2, they also used inconsistent conventions for the
traces.  The correct formul\ae\ should read
$$
\eqalignno{
{\bf B}_a(t)\,\becomes{t\to\infty}\,
    -\coeff{11}{3} \tr_{V,M=0}(Q_a^2)\,+\, \coeff23\tr_{F,M=0}(Q_a^2)\,
    +\,\coeff13\tr_{S,M=0}(Q_a^2)\, &
\equiv\,b_a\,, &(6)\cr
{\cal B}_a\,\becomes{\tau_2\to\infty}\,
    -\coeff{11}{3} \tr_{V,M=0}(Q_a^2)\,+\, \coeff23\tr_{F,M=0}(Q_a^2)\,
    +\,\coeff13\tr_{S,M=0}(Q_a^2)\, &
=\, b_a\, \equiv\,\lim_{t\to\infty} {\bf B}_a(t) ,\qquad& (23)\cr }
$$
where the traces are taken over the massless charged particles
{\sl and count each CPT-conjugate particle-antiparticle pair only once}.

\par\smallskip
%\noindent {\bf($\bullet$)}\enspace
Most of the above errors were either obvious or irrelevant to most of
the readers of the paper.
Unfortunately, the missing factor of 2 in eqs.~(22) and (23) was
neither,
which lead to its propagation via papers concerned with the threshold
corrections in specific string models.
I would like to use this opportunity to apologize to the authors of
those
papers for leading them into the error.
I would also like to apologize for propagating this error myself,
in the article {\it ``Moduli-Dependence of String Loop Corrections
to Gauge Coupling Constants''} I co-authored with L.~Dixon and J.~Louis
({\sl Nuclear Physics \bf B355} (1991), p.~649).
Fortunately, only some of the intermediate results of that article
are affected by this missing factor of 2; the final results --- the
relations
between $\Delta_a$ of an orbifold and the $b'_a$ ($b_a$ coefficients
of the $N=2$ partial orbifold) are correct, provided one uses the
correct
definitions of the $b_a$ and $b'_a$ coefficients.

The author thanks Ignatios Antoniadis, for pointing out the above errors
(even though his attempts to correct them were also erroneous),
and Mirjam Cveti\v{c}, for convincing me that these errors must be
corrected
and for helping me to correct some of them.

\bye